%% file: ms.tex
\documentclass[letterpaper,twocolumn,10pt]{article}
\usepackage{usenix}
\usepackage{authblk}

\usepackage{tikz}
\usepackage{amsmath}

\usepackage[normalem]{ulem} 
\usepackage{enumitem}

\newcommand{\itemph}[1]{\textit{#1}}
\renewcommand{\emph}[1]{\itemph{#1}}

\usepackage{comment}
\usepackage{amsmath,amssymb,amsthm, amsfonts,color}
\newcommand\thankssymb[1]{\textsuperscript{*}}

\newcommand\dmitrii[1]{\noindent{\color{brown} {\bf \fbox{Dmitrii}} {\it#1}}}
\newcommand\plamen[1]{\noindent{\color{green} {\bf \fbox{Plamen}} {\it#1}}}

\newcommand\boris[1]{\noindent{\color{blue} {\bf \fbox{Boris}} {\it#1}}}

\newcommand\changes[1]{\noindent{\color{red} {\bf \fbox{C}} {#1}}}

\usepackage{titling}
\pretitle{\begin{center}\vspace{-.5in}
\textbf{In Proceedings of 14th USENIX Workshop on Offensive Technologies (WOOT'20)}\\
\vspace{.35in}
\normalfont\Large\bfseries}
\posttitle{\par\end{center}
}
\preauthor{\begin{center}\normalfont\large%
    \begin{tabular}[t]{c}}
\postauthor{\end{tabular}\par\end{center}}

\begin{document}

\date{}

\title{\Large \bf Bankrupt Covert Channel:\\
  Turning Network Predictability into Vulnerability}


\author{Dmitrii Ustiugov\thanks{The first two authors contributed equally to this work.} \hspace{0.4in} Plamen Petrov\textsuperscript{\color{green}{*}} \hspace{0.4in} M.R. Siavash Katebzadeh\hspace{0.4in} Boris Grot\\
\textit{University of Edinburgh}
\vspace{-2.5em}
}

\maketitle

\thispagestyle{empty}

\begin{abstract}

Recent years have seen a surge in the number of data leaks despite aggressive information-containment measures deployed by cloud providers.
When attackers acquire sensitive data in a secure cloud environment, covert communication channels are a key tool to exfiltrate the data to the outside world. 
While the bulk of prior work focused on covert channels within a single CPU, they require the spy (transmitter) and the receiver to share the CPU, which might be difficult to achieve in a cloud environment with hundreds or thousands of machines. 

This work presents Bankrupt, a high-rate highly clandestine channel that enables covert communication between the spy and the receiver running on different nodes in an RDMA network. In Bankrupt, the spy communicates with the receiver by issuing RDMA network packets to a private memory region allocated to it on a different machine (an intermediary). The receiver similarly allocates a separate memory region on the same intermediary, also accessed via RDMA. 
By steering RDMA packets to a specific set of remote memory addresses, the spy causes deep queuing at one memory bank, which is the finest addressable internal unit of main memory. This exposes a timing channel that the receiver can listen on by issuing probe packets to addresses mapped to the same bank but in its own private memory region. 
Bankrupt channel delivers 74Kb/s throughput in CloudLab's public cloud while remaining undetectable to the existing monitoring capabilities, such as CPU and NIC performance counters.

\end{abstract}

\section{Introduction}
\label{sec:intro}
\input{Sections/1-Introduction}


\section{Background}
\label{sec:background}
\input{Sections/2-Background}


\section{Threat Model}
\label{sec:threat}
\input{Sections/3-ThreatModel}

\section{The Bankrupt Channel}
\label{sec:attack_overview}
\input{Sections/4-BankruptChannel}


\section{Methodology}
\label{sec:method}
\input{Sections/5-Methodology}

\section{Evaluation}
\label{sec:results}
\input{Sections/6-Results}

\section{Detection and Mitigation}
\label{sec:detectionMitigation}
\input{Sections/7-DetectionMitigation}
\section{Responsible Disclosure \& Code Availability}
\label{sec:responsible}
\input{Sections/8-ResponsibleDisclosure}

\section{Related Work}
\label{sec:related}
\input{Sections/9-RelatedWork}


\section{Conclusion}
\label{sec:conclusion}
\input{Sections/Conclusion}

\input{Sections/Ack}

\bibliographystyle{plain}
\bibliography{bibfile.bib}


\end{document}

%% file: Sections/1-Introduction.tex
\vspace{-0.4em}

In the digital era, information security has become the crucial necessity that drives private and public cloud vendors to take all available measures to contain sensitive and confidential data within customer and government-defined boundaries. Despite the efforts, security breaches are commonplace, often compromising highly sensitive data including 
personal information~\cite{upguard_cloud_leak,info_age_cloud_leak,easyjet_leak, dropbox_personalData}, passwords~\cite{zdnet_pw,linkedin_pw,cybersec_insiders_cloud_leak}, and medical records~\cite{healthcare1,healthcare2}.

Often staying in the shadow of side channels that acquire the sensitive data, covert communication channels are a critical tool used by attackers to exfiltrate the data from a secure environment. Due to strict information containment measures, like firewalls, a \emph{spy} software may not have a direct access to the Internet and, instead, has to communicate -- via a covert channel -- the acquired data to a co-operative \emph{receiver} software that is outside of the secure environment and with Internet access. Recent study by Reardon et al. spotlights a wide usage of covert channels by attackers in the real world~\cite{reardon201950}.


From an attacker's perspective, constructing an efficient covert channel poses several practical challenges.  
In a public or private cloud with thousands of nodes, it may be difficult to ensure that spy and receiver applications get scheduled to the same node.
A more reliable strategy is to construct a covert channel that works across the datacenter, thus allowing the spy and the receiver to reside on different nodes. 
Second, the channel needs to be fast enough to transmit any amount of sensitive data, which could reach into gigabytes (e.g., medical records~\cite{healthcare2,healthcare1}). Finally, the channel must remain stealthy even if the cloud vendor is actively monitoring for its existence.

In this work, we introduce a timing-based covert communication channel, called Bankrupt, that meets all of these requirements. 
Bankrupt enables high-rate covert communication across a datacenter's RDMA-enabled network, thus decoupling the physical placement of the spy and the receiver from their ability to communicate covertly while avoiding detection by existing monitoring facilities,
including CPU and NIC performance counters.

Bankrupt relies on an RDMA-enabled network, a technology that is being adopted by public and private cloud operators~\cite{azure_RDMA,alibaba_RDMA,huawei_RDMA,guo2016rdma}.
The wide deployment of RDMA technology is due to its fundamental latency advantages over traditional datacenter networks. These advantages come from the fact that RDMA offers direct access to a remote server's memory via an RDMA-enabled NIC with full bypass of the remote CPU. The Bankrupt attack turns the low latency and predictability of RDMA into a vulnerability, as predictable latencies enable a highly robust timing channel. 

\begin{figure}[h!]
\centering
\includegraphics[width=0.96\columnwidth]{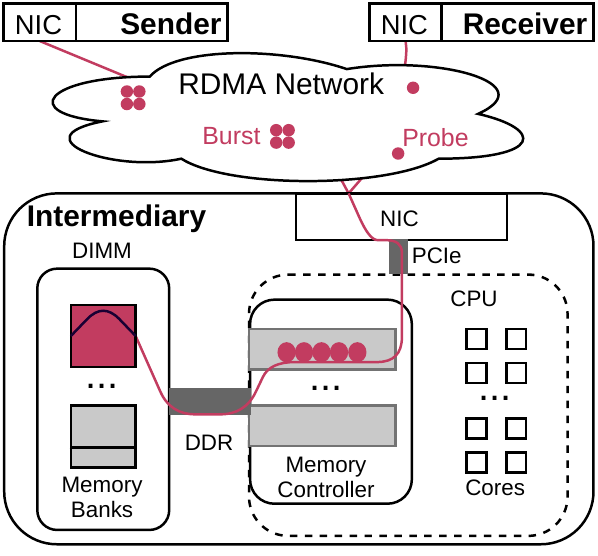}
\vspace{-0.5em}
\caption{Bankrupt attack environment and operation. 
}
\label{overview}
\end{figure}

Bankrupt uses a remote node's main memory as a timing channel accessed via RDMA. 
Figure~\ref{overview} shows how the attack works. In the figure, the spy (also referred to as a sender) wishes to communicate a secret to the receiver residing on a different node. Although unable to communicate directly, e.g., in case of logical separation of the network,
both the sender and the receiver have access to another node's (an intermediary) memory (e.g., a storage server~\cite{smb_direct,nfs_rdma}), where both the sender and the receiver have legitimately allocated private RDMA-exposed memory regions. The sender and the receiver do \emph{not} share physical memory, thus being isolated from the cloud operator's perspective.
The sender sends a stream of RDMA read requests to its private memory region at the intermediary, but concentrates all of the reads on a single memory bank inside the intermediary's memory.\footnote{A memory bank is a memory-internal device that hosts the data. A modern memory subsystem comprises up to hundreds of banks~(\S\ref{sec:memory101}).} 
By exploiting the much higher bandwidth of RDMA (up to 200Gb/s with today's commodity NICs) in comparison to that of a single memory bank (\textasciitilde10Gb/s), the spy can trivially induce queuing at the target bank, causing the latency of accesses to that bank to spike. 
By issuing RDMA reads to the same bank, the receiver can detect the latency spike. The presence or absence of high latency, as in a modulated analog signal, informs the receiver of the current bit's value. 


In order for Bankrupt to succeed, both the spy and the receiver must be able to consistently target a single memory bank on a server with hundreds of banks. A memory controller inside the intermediary CPU uses a subset of physical address bits (the bank bits) to route the memory requests, originated both from the CPU and the RDMA NIC, to the target bank. As RDMA legitimately exposes the virtual addresses of the intermediary, we show that a remote attacker is able to use this information to reverse-engineer the position of the bank bits in the address space of the intermediary with an 
algorithm that is of linear complexity in the number of address bits.


Another challenge for Bankrupt is to guarantee that the latency spike on the target node is sufficiently high that the receiver on another node can reliably observe it by issuing probes over RDMA to the same bank. 
The quality of the signal may be compromised by network noise or the presence of memory-intensive applications on the intermediary. However, our experiments show that the massive bandwidth offered by today's RDMA networks and memory subsystems create a favorable environment for a highly robust channel. 

We evaluate the Bankrupt channel in a private cluster and on CloudLab, a large-scale public cloud used by computer researchers. In the private cluster, the Bankrupt channel achieves 
114Kb/s throughput in isolation. Under network load, 
Bankrupt still provides 67Kb/s throughput, as the sender needs to issue larger bursts of RDMA reads to overcome the noise level. We find that the channel's bandwidth and accuracy are unaffected in the presence of local memory traffic from innocuous memory-intensive applications that concurrently run on the intermediary because a CPU efficiently balances regular memory traffic. On CloudLab, we show that Bankrupt delivers 74Kb/s channel throughput.


To summarize, our contributions are as follows:
\vspace{-0.5em}

\begin{itemize}[leftmargin=*]
    \item We construct a timing channel, called Bankrupt, that enables covert communication between participants, which share the same RDMA network but run on different nodes, via queuing inside remote memory hosted on a third -- innocuous -- node. 
\vspace{-0.6em}
    \item We provide guidelines for setting up Bankrupt with an arbitrary RDMA network and CPU; namely for discovering the addresses that belong to the target memory bank in remote memory as well as establishing and dynamically adjusting the channel's modulated signal by issuing shaped bursts of RDMA traffic even in a noisy environment.
\vspace{-0.6em}
    \item We show that the attack is undetectable by software-based memory access latency monitoring, Infiniband NIC and CPU performance counters. 

\end{itemize}

%% file: Sections/2-Background.tex
The Bankrupt channel leverages features of a modern RDMA-enabled network and the CPU memory subsystem. 
This section introduces the relevant aspects of these technologies in order to explain how they work together in Bankrupt.
\vspace{-0.6em}

\subsection{Covert Channels}
\label{sec:covert101}
\vspace{-0.4em}

Covert channels, introduced first by Lampson~\cite{lampson1973note}, enable communication between independent entities over isolation boundaries, bypassing firewalls and communication auditing. The basic setup of a covert channel consists of the two participants: a spy (or sender) program that would like to communicate (exfiltrate) the previously obtained secret to the outside world, and a receiver that is aware of the sender's existence. Finally, the sender 
and the receiver should share a common resource that they can use to communicate and decode the signal.

To communicate, the sender and the receiver usually exploit resource sharing in a computer system by establishing a communication channel via legitimate innocuous actions. For example, a timing channel can be established by modulating access latencies to a shared CPU cache~\cite{Maurice2015C5CC, yarom2014flush+, liu2015last}, main memory~\cite{pessl2016drama, xiao2012covert, xiao2013security} or by modulating temperature on a multi-core CPU chip~\cite{masti2015thermal}. 

Despite the variety of the previously proposed covert channels, only a few of them may pose a threat in a realistic deployment scenario, as indicated by prior work~\cite{wu2014whispers}. We identify three key requirements that a practical covert channel has to meet for maximum impact and generality, namely \emph{wide visibility}, \emph{high communication bandwidth}, and \emph{stealthiness}. We next discuss each of these requirements. 

{\bf Wide visibility:} 
An attacker seeking to exfiltrate secrets needs a covert channel that avoids or minimizes the degree of colocation between the sender and the receiver. 
While the most straight-forward way for the sender and the receiver to covertly communicate is by co-running on the same node,
such a high degree of colocation is difficult to achieve in practice because cloud vendors purposefully randomize placement of the applications on their nodes. However, many providers do provide their customers with means to localize applications within the same physical network to offer low latency networking~\cite{aws:placement,azure:placement}. Hence, a cross-network covert channel may significantly simplify the attack setup for an attacker.

{\bf High bandwidth:} Certain types of sensitive data may have a significant volume, potentially reaching into gigabytes (e.g., healthcare records~\cite{healthcare2}). Hence, to be broadly applicable, a covert channel should deliver high transmission bandwidth channel to exfiltrate data of an arbitrary size. Moreover, a channel's  bandwidth should remain high even in the presence of various types of system noise, e.g., network traffic or CPU activity of numerous innocuous applications and system services that share the same network or CPU.

{\bf Stealthiness:}
An ideal covert channel should remain stealthy even if platform owners suspect its existence and actively monitor for it. Indeed, modern servers feature a wide range of monitoring capabilities, for example, CPU performance counters and network latency anomaly detectors~\cite{perfTool, Zhang2016, Chen2018, cziva_ruru, Joshi, Popescu2017a, popescu18_dcn_ptpmes}. A truly covert channel should remain undetectable by these techniques. 

Previously proposed covert channels struggle to comply with all three 
requirements. CPU-cache based covert channels 
restrict the sender
and receiver colocation scenario to the same physical core or CPU chip~\cite{Maurice2015C5CC, yarom2014flush+, liu2015last, wu2014whispers}.
Cache-based channels also tend to increase the miss rate of the caches exposing themselves to system monitoring software~\cite{gruss2016flush+,wu2014whispers}.
The attacks across a network often struggle to deliver bandwidth higher than 1Kb/s~\cite{cross-router,tahir2016sneak} or may be difficult to use in a noisy environment due to a relatively small latency gap in their timing channel~\cite{kurth2020netcat}.

\vspace{-0.4em}

\subsection{RDMA Networks}
\label{sec:rdma101}
\vspace{-0.4em}

Originating from high-performance computing, Remote Direct Memory Access (RDMA)  networks see rapid adoption by cloud providers that strive to deliver low latency and high throughput to their customers at the scale of an entire datacenter~\cite{azure_RDMA,alibaba_RDMA,huawei_RDMA,guo2016rdma}. 
For example, Azure customers can rent a virtual RDMA-connected cluster of 80 000 virtual cores~\cite{microsoftRDMA200g}.

RDMA allows the applications to access the remote memory via user-level \emph{one-sided} RDMA read and write primitives that bypass the destination CPU completely, allowing network packets to directly read or write to its memory.
First, to expose remote memory for RDMA operations, an application running on one node needs to register a memory region on the node that hosts the memory. 
After that, the application can issue RDMA read (write) network packets to read (write) the remote memory contents directly -- by specifying the remote node's virtual address and the accessed memory chunk length. 


When an RDMA packet arrives at the destination node's 
NIC, the NIC translates the virtual address, specified in the packet's header, to the corresponding physical memory address and engages a DMA (read or write) transaction. The root complex unrolls the transaction into a sequence of CPU cache-block sized memory requests that eventually reach a memory controller of the CPU. Finally, the memory controller serves these requests from the destination's main memory and sends the responses back to the NIC. Upon receiving the responses from the memory controller, the NIC forms a response packet and sends it out to the requesting node.


\begin{figure}[h!]
\centering
\includegraphics[width=0.97\columnwidth]{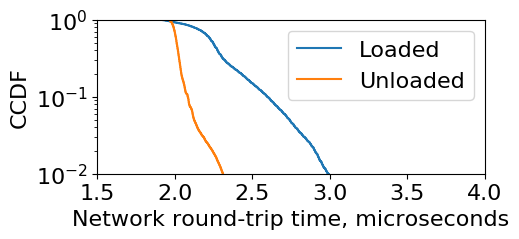}
\vspace{-0.8em}
\caption{Round-trip latency, as a complementary cumulative distribution function (CCDF), of an RDMA read operation in isolation and when network links and switch run at 70\% utilization. See \S\ref{sec:method} for the setup details.
}
\label{rdma_ccdf}
\end{figure}

Because RDMA bypasses the destination CPU, the network round-trip time remains predictable even in the presence of other network-intensive applications that use the network infrastructure simultaneously. To showcase RDMA network predictability, we plot the 64B RDMA read latency, in isolation and under network load of \textasciitilde40Gb/s (which represents \textasciitilde70\% of the link bandwidth) from a client node to a server node (Figure~\ref{rdma_ccdf}). 
Despite the loaded network links and switch ports, the $99^{th}$ percentile of the network round-trip delay is larger than in the unloaded case by just 1.2 microseconds.
This high RDMA predictability indicates the potential for constructing a robust timing channel in an RDMA network.

\vspace{-0.4em}

\subsection{Memory Organization}
\label{sec:memory101}
\vspace{-0.4em}


The memory subsystem of a modern server CPU has a hierarchical structure:
there are 2-6 memory channels, each with its own memory controller, and each channel has 2-4 DIMMs. Each DIMM consists of up to 64 independent memory devices called memory banks.\footnote{DIMMs normally consist of 1-4 ranks, and each rank of 8-16 banks.} In total, today's server may have hundreds of memory banks across its multitude of DIMMs, channels and sockets~\cite{amd_rome}. 

The high overall bandwidth of a modern memory system is delivered by this ensemble of small independent banks, where each bank provides only a small fraction of the total memory bandwidth.
Internally, a memory bank stores data as an array of DRAM rows, bringing one row at a time to an SRAM row buffer before serving a memory request. 
A single bank's bandwidth is bounded by the time the bank takes to bring and serve the corresponding DRAM row from the array. Assuming no access locality, this time is the sum of the three key DRAM latency components, namely $tCL$, $tRCD$, and $tRP$. Each of the three latency components is 13-15ns which means that the total latency of serving a single 64-byte memory request is 39-45ns, resulting in the peak theoretical 1.4-1.6GB/s bandwidth of a single bank. Due to the end of DRAM technology scaling, these key latency components have stayed the same over the last two decades~\cite{ddr2Datasheet,ddr4Datasheet}.

The memory controller manages each bank independently from others, hence all requests to a particular bank reside in a dedicated per-bank queue inside the memory controller. The memory controller routes memory requests to banks based on their physical addresses by employing a manufacturer-specific fixed hash function (referred to as a memory interleaving scheme) that determines the destination bank based on a subset of the memory request's physical address.

To deliver maximum overall memory bandwidth, manufacturers choose the scheme that maximises the parallelism across memory banks. For example, many processors use a cache-block interleaving scheme that implies that accesses to adjacent cache blocks are served from different banks. While some manufacturers disclose memory interleaving information, others keep this information proprietary~\cite{pessl2016drama,amd:interleaving}.

\vspace{-0.4em}

%% file: Sections/3-ThreatModel.tex

\vspace{-0.4em}

We assume a highly restricted setting that is similar to the public cloud environment~\cite{azure_RDMA,alibaba_RDMA,huawei_RDMA,guo2016rdma}.
Both sender and receiver run either natively or in a virtual machine. They do not have means to guarantee colocation on the same server, and thus, might be unable to leverage one of the known local (i.e., on the same node) covert channels. Furthermore, the sender and receiver are not allowed to directly communicate with each other over the network, as if the network is logically separated.
We assume that no privilege escalation is possible and both the sender's and the receiver's applications have normal user privileges. The applications are unable to alter any of the existing firewall policies and do not have the capability to observe the network traffic.

To establish a covert communication channel in this environment, the sender and the receiver can instead communicate via an intra-datacenter network by modulating local or remote memory access latency. For example, both the sender and the receiver may be able to allocate non-shared remote memory regions on one of the shared storage servers~\cite{smb_direct,nfs_rdma}, as assumed by prior work~\cite{kurth2020netcat}.
The remote memory access latency can be modulated by performing one-sided read and/or write operations via RDMA. Ideally,  the sender can allocate remote memory on a receiver's node, allowing the receiver to observe memory access latency on its own server. However, this might be difficult to guarantee as the sender has no knowledge where the receiver is running, and furthermore, is unable to control where its remote memory is allocated by the datacenter resource manager. 

With Bankrupt, the sender and the receiver can communicate by using the remote memory of one of the other nodes in the cluster (further referred to as an intermediary), where both sides can allocate remote memory. 
To find an intermediary, both the sender and the receiver may request many storage servers and periodically create and search for the modulated signal inside the remote memory of each server
by periodically probing a specific set of addresses (\S\ref{sec:addresses}) in each server's memory with RDMA reads.

The intermediary is non-malicious and the attacker does not have direct control over it. 
The sole requirement is that the sender and the receiver need to have access to their corresponding -- completely private -- memory regions that have been allocated for their exclusive use with RDMA one-sided operations (i.e., remote reads and writes). 

\vspace{-0.6em}

%% file: Sections/4-BankruptChannel.tex
\vspace{-0.4em}

The Bankrupt attack turns the main advantage of RDMA networks -- the direct access to remote memory -- into a high-rate stealthy communication channel that remains robust even in the presence of noise arising from the network and other innocuous co-running applications/VMs. 

In this section, we describe the attack, introducing a cross-network timing channel,
and provide guidelines for setting up the Bankrupt channel in an arbitrary RDMA cluster. 

\vspace{-0.4em}
\subsection{The Bankrupt Timing Channel}
\vspace{-0.4em}

Direct access to remote memory enables the sender to create a fast timing channel by modulating the access latency to a single memory bank by selectively steering legitimate network traffic to a small subset of memory addresses in the intermediary's memory. 






The timing channel relies on several features of the RDMA NIC and the CPU in a modern server.
First, CPU and RDMA network cards are tightly integrated via PCIe. This allows the network traffic to flow directly into the destination CPU's memory subsystem without any software mediation from the destination CPU (\S\ref{sec:rdma101}). Second, the memory bank addressing (memory interleaving) scheme is static, which enables a remote attacker to steer their network traffic to a single memory bank (\S\ref{sec:memory101}). 
Finally, the bandwidth disparity between the 100-200Gb/s network and \textasciitilde10Gb/s memory bank enables a remote attacker to easily modulate the access latency of a single memory bank by forcing queuing at the target memory bank with excessive traffic.

To set up reliable communication, we use a unidirectional communication protocol. The sender transmits information as a synchronous analog signal by modulating, at a fixed frequency,
the response latency of a single bank that is located in remote memory.
By doing that, the sender modulates the entire network round-trip delay, as observed by the receiver, for those RDMA packets that go to the target bank in the remote memory.
To transmit a \texttt{1}, 
the sender switches the target memory bank to the \emph{contended} state (i.e., high access time during a period) by steering the network traffic in the form of bursts of RDMA read operations to the bank.\footnote{Cache blocks, accessed by RDMA reads, are not allocated in a CPU's last-level cache to avoid cache pollution~\cite{directtechnology}.} To transmit a \texttt{0}, the sender does not issue network traffic, leaving the bank in the \emph{uncontended} state (i.e., low access time) for the time period that is defined by the sending frequency. 

To read the transmitted signal, the receiver issues probes, which are normal RDMA read operations, at a certain frequency to measure and record the observed remote memory access latency. The sender splinters the messages into fixed-sized packets with a pre-defined preamble as a packet header, followed by a payload. The preamble contains a number of bits that would indicate the beginning of a message to the receiver while the payload contains the transmitted information.
Transmission accuracy can be further improved by adding error correction codes to the packet header, though we do not take advantage of it in this work.

\vspace{-0.4em}
\subsection{Bankrupt Setup Guidelines}
\vspace{-0.4em}


To set up the channel, both the sender and the receiver need to find a subset of remote memory addresses that belong to the same memory bank in the intermediary's memory -- this bank will serve as the actual timing channel.
The sender needs to adjust the size of the bursts of RDMA operations and to select the appropriate sending frequency. The receiver needs to tune its probing frequency. We next discuss each of these requirements.

\vspace{-0.4em}
\subsubsection{Identifying RDMA Addresses that Belong to the Same Memory Bank}
\label{sec:addresses}

The sender and the receiver (which, together, we refer to as the ``attacker'') can use an arbitrary bank in the intermediary. The choice of the bank is defined by the subset of the virtual address bits (further referred to as \emph{bank bits}), the value of which tells the memory controller which bank to route a memory request to (\S\ref{sec:memory101}). For simplicity, the attacker can set the bank bits to a pre-defined value (e.g., all zeros). The attacker (both the sender and the receiver can do it separately) just needs to find any addresses with bank bits set to that value.

The attacker starts by locating the bank bits. These bits are normally located in the least-significant, i.e., [6:$X$] where $X>6$, part of the address 
in order to balance the load across all available banks. Hence, the attacker needs to derive $X$ to locate the positions of the bank bits.

Since modern computer systems manage memory at the granularity of pages (Linux/x86 uses 4KB, 2MB, and 1GB pages), a subset of least-significant bits (LSBs) in a virtual address and the corresponding physical address are identical. 
In RDMA-connected systems where low latency is a priority, the vendors often use large, 2MB or 1GB, pages for remote memory to avoid the bulk of translation cache misses in RDMA NICs~\cite{dragojevic:fast,Cong2018,wang18_rdmav,novakovic19_storm}. If 2MB or 1GB pages are used, virtual and physical addresses share 21 or 30 LSBs, respectively. 
Knowing 30 bits is enough to locate the positions of all the bits in a virtual address that identify a bank for all the tested systems both in this work and in prior work~\cite{pessl2016drama}. For some systems, knowing 21 LSB bits  suffices. 
The attacker can identify the positions of the bank bits in an address by using only RDMA read operations in a search algorithm, which is of linear complexity in the number of address bits.
This algorithm allows to find addresses that belong to a single bank by iteratively excluding addresses belonging to other banks.

First, the attacker chooses a set of unique arbitrary virtual addresses that are 64-byte (cache-block) aligned, i.e., the bits [0:5]  in the address must be zero while bits [6:63] have arbitrary values. The number of addresses needs to be big enough to not fit in a memory controller's hardware buffer at once to avoid coalescing requests to the same address. We find that 64 addresses are enough for all the platforms that we tested.

Second, the attacker issues RDMA reads to those addresses and estimates the resulting network throughput based on the number of RDMA reads that complete in the measurement interval. 
If the traffic exceeds the throughput of a single memory bank (\textasciitilde1.2GB/s in our experiments), the attacker infers that memory requests are being served by several banks. 
To exclude addresses from a half of the banks in each iteration, the attacker sets one more LSB bit to 0 ([0:6] bits in the second iteration, [0:7] in the third one, etc.) and estimates the network throughput. 
After a number of iterations, the throughput converges to that of a single bank, meaning that the most-significant bank bit is set to zero, revealing the $X$ parameter. Note, that not all bits in [6:$X$] define a bank so some iterations may not lead to a throughput reduction.

\vspace{-0.4em}
\subsubsection{Tuning Burst Characteristics}
\label{sec:burst_characteristics}
\vspace{-0.4em}


The sender modulates the access time of the target memory bank by forcing queue build-up at the target bank's queue inside the memory controller. 
To force queuing, the sender issues bursts of RDMA reads to addresses that map to the target bank. 
Below we provide guidelines for the sender to shape the traffic to produce a robust signal that is visible across the RDMA network.

To guarantee visibility of the signal, the sender must issue bursts that require a large amount of time to drain by the target bank (e.g., a microsecond or more). A memory bank serves read accesses to any of these addresses serially and each one within a fixed service time that is in the 39-45ns range~(\S\ref{sec:memory101}). This enables the 
sender
to accurately estimate the service time of a bank depending on the queue depth. Given the predictability of the RDMA network~(\S\ref{sec:rdma101}), it is possible to accurately modulate the RDMA network round-trip delay.

\vspace{-0.4em}
\subsubsection{Tuning Sender's and Receiver's Frequency}
\label{sec:tuning}
\vspace{-0.4em}
To maximize the transmission bandwidth of the channel, the sender needs to find the maximum frequency, i.e., the minimum period, which depends on the time the target bank takes to serve a burst as well as the level of noise in the system. 
There are two types of noise that may impact Bankrupt's transmission speed and accuracy. First is the network traffic from other innocuous applications, which share the same physical network, that may increase network delays. The other type of noise is the memory traffic that comes from innocuous software that runs on the intermediary, causing extra queuing at the memory banks.

To overcome the noise, the sender may need to increase the burst size so that the modulated delay is higher than the noise level, i.e., maintain the signal/noise ratio more than 1. Hence, the sender should periodically adjust its frequency according to the environment changes: larger bursts improve the signal/noise ratio but take longer to drain which limits the sending frequency. To determine the optimal sending frequency, both the sender and the receiver measure the current network round-trip time (which serves as an indicator of the noise level) and determine the minimal required burst size for the current noise level. Given the pre-defined preamble in packet headers, the receiver can derive the sending frequency and adjust to its changes.




The receiver needs to issue probes, in the form of RDMA reads, to the intermediary's bank to detect the transmitted signal. The probing frequency needs to be high enough to allow the receiver to decode the bit transmitted in each sending period. Before the transmission starts, the receiver first determines the unloaded latency as the 95$^{th}$ percentile latency of the network round-trip when the channel is inactive, to discount the rare RDMA network latency outliers. 
Then, during each sending period, the receiver performs several measurements of round-trip time and compares the round-trip latency to the unloaded latency to determine if the target bank is in the contended state. The receiver computes the round-trip latency as the arithmetic mean of the measurements that are taken in the middle of that period to capture the peak of the potential bank contention, by dropping first and last 25\% of the measurements. If the round-trip latency is larger than the unloaded latency, the receiver records \texttt{1}, otherwise a \texttt{0}. To account for sporadic changes in network load, the receiver periodically re-computes the unloaded latency.

\vspace{-0.6em}

%% file: Sections/5-Methodology.tex
\vspace{-0.4em}

We evaluate Bankrupt in terms of transmission bandwidth, accuracy, and stealthiness.

We mount Bankrupt on a private cluster to examine the performance in isolation and under various types of load. To demonstrate the feasibility of the Bankrupt channel in a public cloud, we evaluate it on CloudLab~\cite{ricci2014introducing,cloudlab}~-- a public cloud in use by Computer Science researchers. 
We evaluated the Bankrupt channel using CloudLab's Utah cluster that connects 200 nodes with a high-speed RDMA network. At the time of testing, the utilization of the nodes in the cluster was 80\%. The specifications of our private cluster and the CloudLab cluster are presented in Table~\ref{table:specs}.
Similarly to prior work~\cite{pessl2016drama}, we enable 1GB large pages on our experimental platforms as they are widely used in RDMA deployments~\cite{novakovic19_storm,wang18_rdmav, Cong2018}. In all of our experiments, the sender and the receiver use RDMA reads of 64 Bytes.





\begin{table}[t!]
\begin{tabular}{c|l|l|}
\cline{2-3}
                                                                              & \multicolumn{1}{c|}{\textbf{\begin{tabular}[c]{@{}c@{}}Private Cluster\end{tabular}}} & \multicolumn{1}{c|}{\textbf{CloudLab Utah}}                                           \\ \hline
\multicolumn{1}{|c|}{CPU}                                                     & \begin{tabular}[c]{@{}l@{}}Xeon E5-2630v4\\ @2.20GHz\\ (Broadwell)\end{tabular}           & \begin{tabular}[c]{@{}l@{}}Xeon E5-2640v4\\ @2.40GHz\\ (Broadwell)\end{tabular} \\ \hline
\multicolumn{1}{|c|}{RAM}                                                     & \begin{tabular}[c]{@{}l@{}}4$\times$16GB, DDR4-2400\end{tabular}                         & \begin{tabular}[c]{@{}l@{}}4$\times$16GB, DDR4-2400\end{tabular}                    \\ \hline
\multicolumn{1}{|c|}{NIC}                                                     & \begin{tabular}[c]{@{}l@{}}Mellanox MT27800\\ CX-5, 56Gb/s\end{tabular}                     & \begin{tabular}[c]{@{}l@{}}Mellanox MT27710\\CX-4 Lx, 50Gb/s\end{tabular}     \\ \hline
\multicolumn{1}{|c|}{\begin{tabular}[c]{@{}c@{}}Core\\ switches\end{tabular}} & ---                                                                      & \begin{tabular}[c]{@{}l@{}}1 Mellanox 2700\end{tabular}                     \\ \hline
\multicolumn{1}{|c|}{\begin{tabular}[c]{@{}c@{}}ToR\\ switches\end{tabular}}  & \begin{tabular}[c]{@{}l@{}}Mellanox SX6012\end{tabular}                               & \begin{tabular}[c]{@{}l@{}}5 Mellanox 2410\end{tabular}                     \\ \hline
\multicolumn{1}{|c|}{OS}                                                      & Ubuntu 18.04                                                                            & Ubuntu 16.04                                                                     \\ \hline
\multicolumn{1}{|c|}{Kernel}                                                  & 4.15.0-58-generic                                                                       & 4.15.0-88-generic                                                                \\ \hline
\multicolumn{1}{|c|}{Nodes}                                                   & 6                                                                                       & 200                                                                              \\ \hline
\end{tabular}
\caption{Specifications of studied platforms.}
\vspace{-1.4em}
\label{table:specs}
\end{table}

To evaluate the bandwidth and accuracy of transmission over the Bankrupt channel, 
we send a sequence of packets where each packet carries a constant preamble and a randomly-generated payload.
For all experiments, unless stated otherwise, we use a 32-bit long preamble of \texttt{10..10} to tune the decoder before decoding the payload of 200 bits. 
We report channel throughput as its {\em true channel capacity}, which we define as achieved channel bandwidth discounting preambles and incorrectly transmitted bits. We measure accuracy as the fraction of correctly decoded bits in the message that we are sending. We keep the receiver's probing frequency set to 500ns in all our experiments as we found that its impact on accuracy is rather moderate.

\vspace{-0.6em}

%% file: Sections/6-Results.tex
\vspace{-0.4em}

In this section we demonstrate the properties of the Bankrupt attack in isolation and under various types of load. We conclude the section with an evaluation of the stealthiness characteristics of the channel and its performance in a realistic datacenter environment.

\vspace{-0.7em}
\subsection{Performance in Isolation} \label{isolation_performance}
\label{sec:isolation}
\vspace{-0.4em}
Following the algorithm in \S\ref{sec:addresses}, we found that the bank bits are located at the positions of bits [6:26] in the address.

\begin{figure}[t!]
\centering
\begin{subfigure}{0.47\textwidth}
    \includegraphics[width=\textwidth]{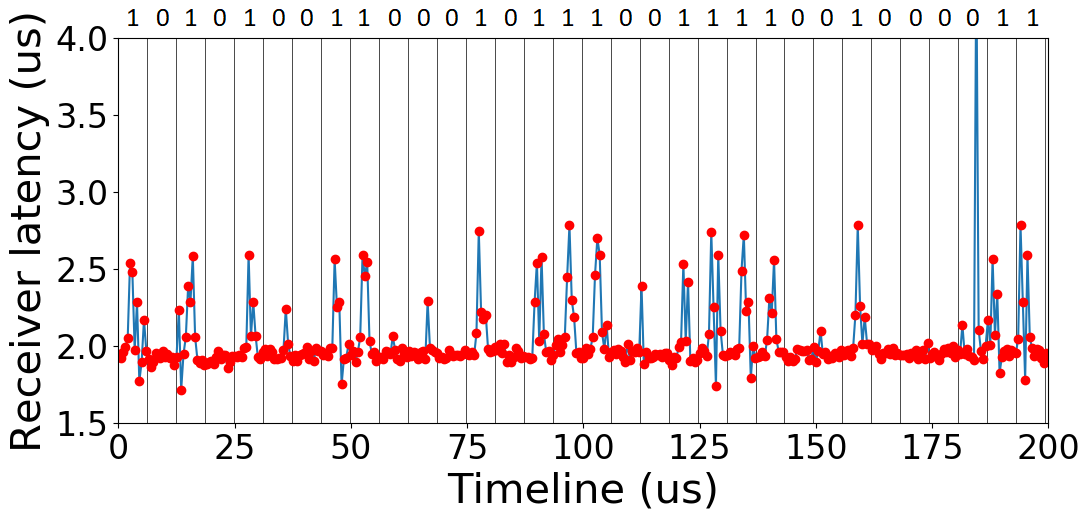}
    \subcaption{Burst size 32}%
\end{subfigure}
\begin{subfigure}{0.47\textwidth}
    \includegraphics[width=\textwidth]{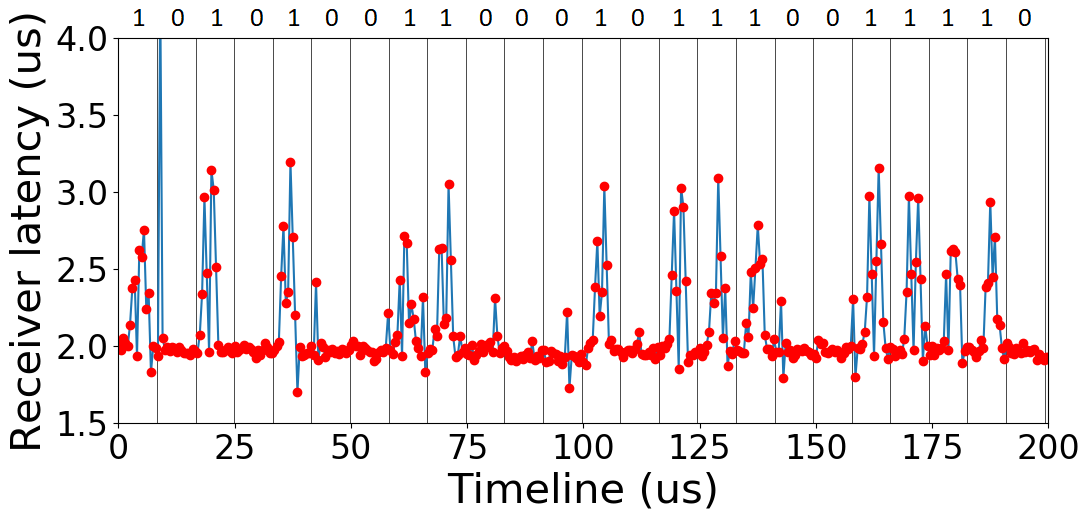}
    \subcaption{Burst size 64}%
    \label{fig:throughput}
\end{subfigure}
\begin{subfigure}{0.47\textwidth}
    \includegraphics[width=\textwidth]{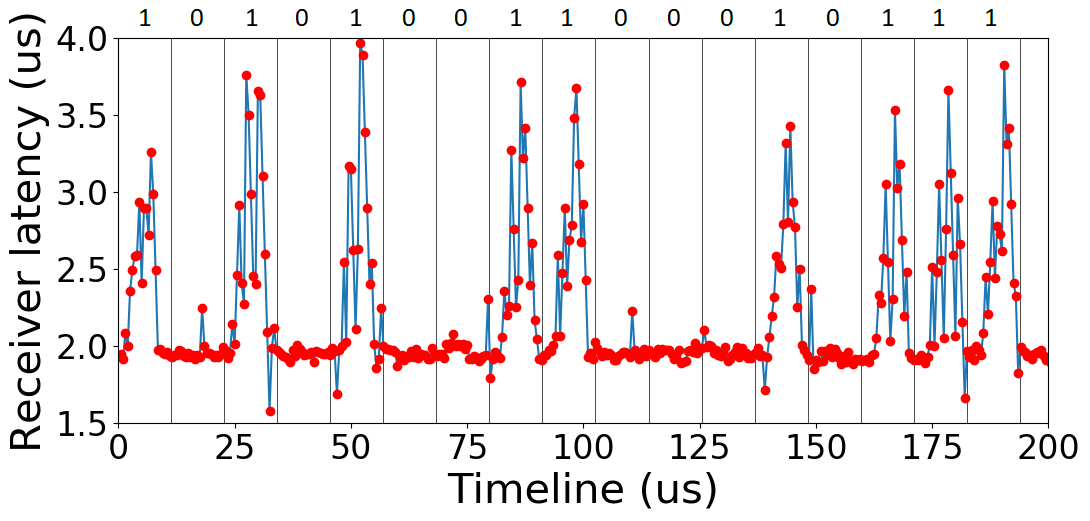}
    \subcaption{Burst size 128}%
    \label{fig:accuracy}
\end{subfigure}
\vspace{-0.4em}
\caption{Signal observed by the receiver in isolation in the private cluster, with the corresponding transmitted messages.}
\vspace{-1.3em}
\label{isolation_cuts}
\end{figure}

Figure~\ref{isolation_cuts} shows that the latency measurements during the transmission for different burst sizes. 
Note that each subfigure shows a different number of transmitted bits in a fixed amount of time (200 microseconds) because varying the burst size also changes the (maximum) sending frequency (\S\ref{sec:tuning}).
The difference between latencies, further referred to as the \textit{latency gap}, that are measured during transmitting \texttt{1}-s and \texttt{0}-s grows with the burst size. 
The minimal burst size that allows decoding is 32 (2KB) that shows the latency gap of 0.5 microsecond.
With the burst size of 128 (8KB), the latency gap is more than twice bigger, around a microsecond that is comparable to the network round-trip time, making the signal more pronounced.




Figure~\ref{bw_acc} shows how the channel throughput and transmission accuracy depend on the burst size. Increasing the burst size from 16 to 32 nearly doubles the transmission accuracy that leads to a surge in the channel throughput, despite the lower signaling frequency with a larger burst size. However, increasing the burst size beyond 32 results in a diminished channel throughput as large burst sizes decrease the signaling frequency that is not compensated for by the gain in accuracy. Thus, we find that the burst size of 32 delivers the highest channel throughput of 114Kb/s (with transmission accuracy of 82.4\%) in isolation in the private cluster.

\vspace{-0.4em}
\subsection{Performance Under Load}
\label{sec:noise}
\vspace{-0.4em}
To evaluate the channel in the presence of noise, we model both the noise coming from other workloads scheduled on the intermediary's CPU and network traffic.

\vspace{-0.8em}
\subsubsection{Local Load}
\label{sec:local_load}
\vspace{-0.4em}

To model the local load in the intermediary's memory, we launch 16 instances of the \texttt{mcf} benchmark, taken from the popular SPEC 2006 benchmark suite ~\cite{henning2006spec}, on the intermediary node. \texttt{mcf} is the most memory-intensive benchmark from that suite~\cite{jaleel2010memory}.
16 instances of \texttt{mcf} that together generate a variable load on the intermediary's memory ranging from 2GB/s to 8GB/s. This is equal to 32MB/s to 128MB/s for each bank, assuming that on average the load spreads evenly among banks.

\begin{figure}[t!]
\centering
\begin{subfigure}{0.47\textwidth}
    \includegraphics[width=\textwidth]{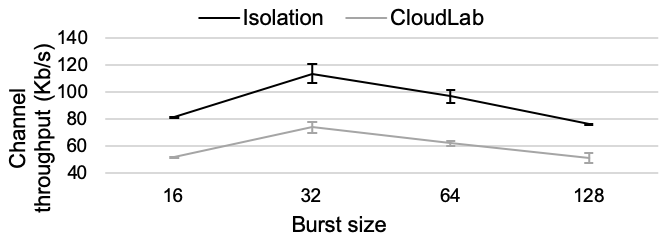}
    \subcaption{Channel throughput as true channel capacity}%
\end{subfigure}
\begin{subfigure}{0.47\textwidth}
    \includegraphics[width=\textwidth]{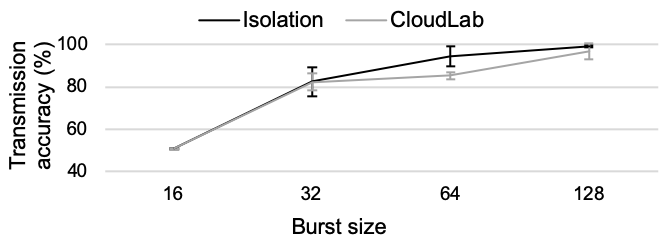}
    \subcaption{Transmission accuracy}%
\end{subfigure}
\vspace{-0.4em}
\caption{True channel capacity and transmission accuracy of the Bankrupt covert channel in isolation and in CloudLab. Error bars show the standard deviation.}
\vspace{-0.8em}
\label{bw_acc}
\end{figure}

\begin{figure}[t!]
\centering
\includegraphics[width=0.47\textwidth]{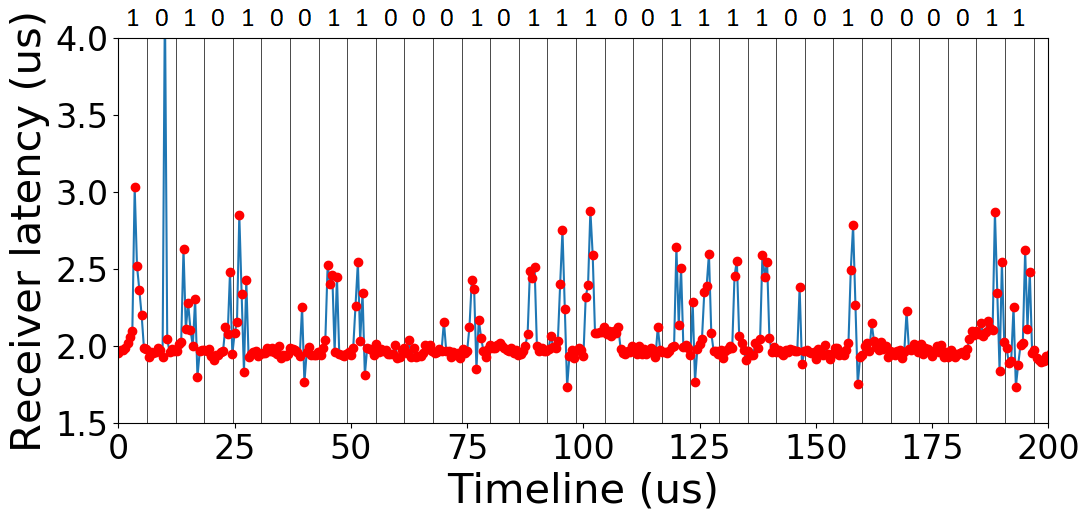}
\caption{Signal observed by the receiver in the presence of local load in the private cluster. The burst size is 32.}
\vspace{-1.4em}
\label{local_load_cuts}
\end{figure}

Figure~\ref{local_load_cuts} shows the signal observed by the receiver for the burst size of 32 in the presence of the local load. The signal is indistinguishable from the one observed in isolation for the same burst size (Figure~\ref{isolation_cuts}). The throughput and the accuracy remain the same as in isolated execution. The channel is unaffected by other workloads running on the intermediary because each individual bank receives a relatively small amount of traffic, since the load is spread across all of the banks in the system (\S\ref{sec:memory101}).


\vspace{-0.4em}
\subsubsection{Network Load}
\label{sec:net_load}
\vspace{-0.4em}

\begin{figure}[t!]
\centering
\begin{subfigure}{0.47\textwidth}
    \includegraphics[width=\textwidth]{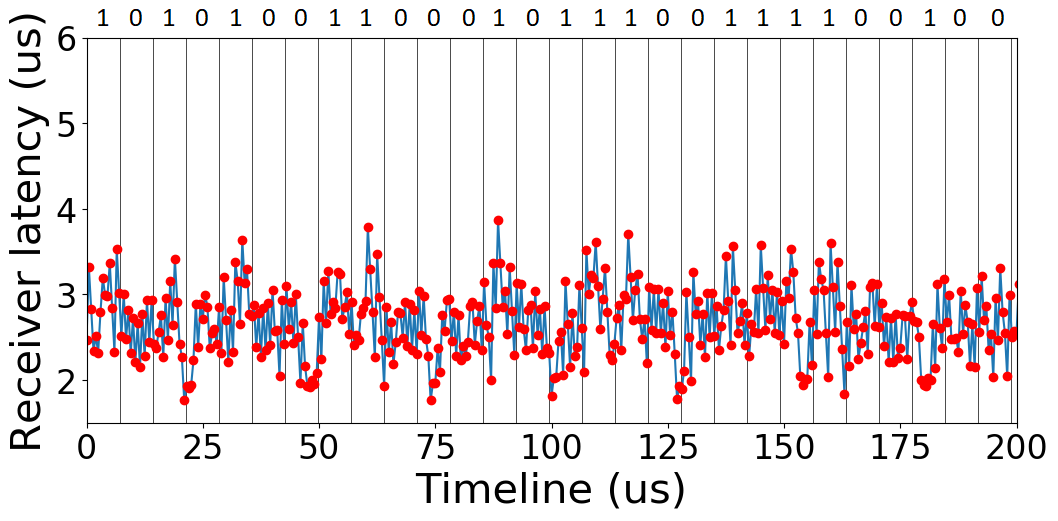}
    \subcaption{Burst size 32}%
\end{subfigure}
\begin{subfigure}{0.47\textwidth}
    \includegraphics[width=\textwidth]{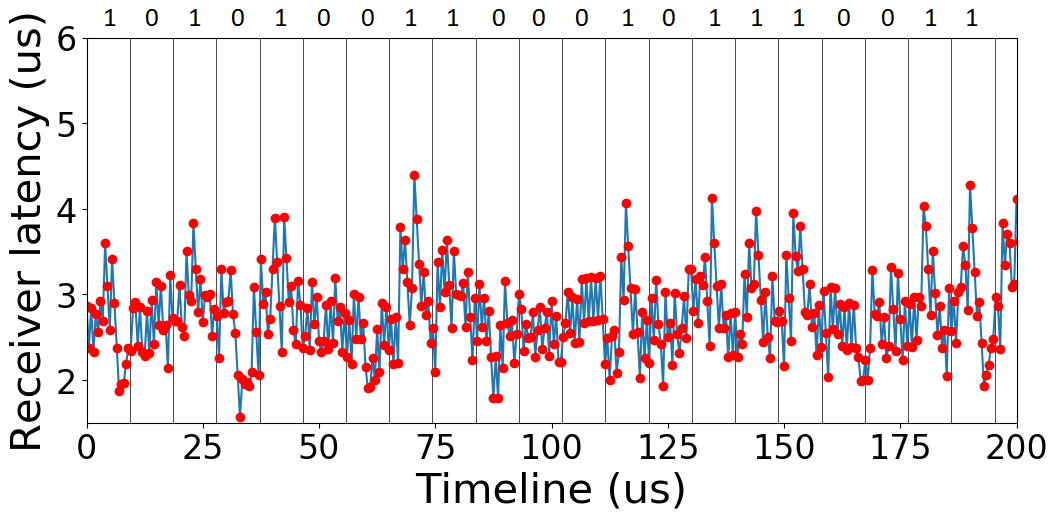}
    \subcaption{Burst size 64}%
\end{subfigure}
\begin{subfigure}{0.47\textwidth}
    \includegraphics[width=\textwidth]{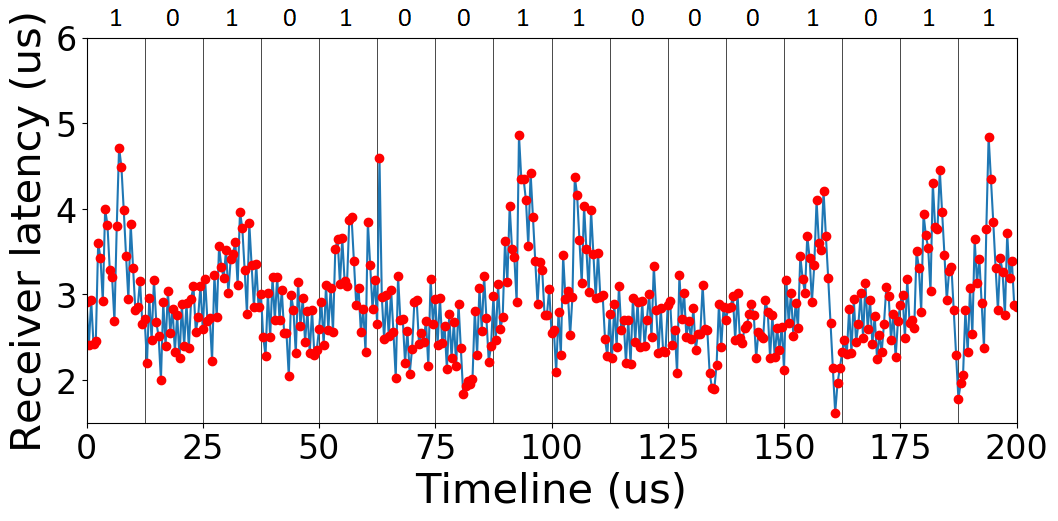}
    \subcaption{Burst size 128}%
\end{subfigure}
\vspace{-0.4em}
\caption{Signal observed by the receiver in the presence of network load in the private cluster.}
\vspace{-1.4em}
\label{net_load_cuts}
\end{figure}

To model network load on the intermediary, we use the \texttt{ib\_read\_bw} benchmark, which is part of the RDMA Perftest \cite{perftest}. We launch \texttt{ib\_read\_bw} from a node, which is different from the three nodes we use for the channel. The network loader generates network traffic to the intermediary, loading the network link, top-of-the-rack switch, and the NIC of the intermediary.
The generated load is equal to \textasciitilde40Gb/s which is around 70\% of the intermediary's total link bandwidth.

Figure~\ref{net_load_cuts} shows the signal observed by the receiver in the presence of network load. With the burst size of 32, the signal is noisy, which prevents its decoding because the round-trip delay, recorded by the receiver, often spikes to 3 microseconds. However, with the burst size of 64-128, the signal is clearly visible atop of the noise as
the latency gap between contended and uncontended states of the bank exceeds 1 microsecond. 

For the burst size of 64, the throughput of the channel reaches 67Kb/s with accuracy of
73.5\%. Compared to the best performance in isolated execution, the throughput decreases by 41\%.
For the burst size of 128, the transmission accuracy is much higher, reaching 95.4\%; however, the resulting channel throughput of 65Kb/s is lower than with a burst size of 64 due to a diminished signaling rate. 

Our experiments suggest that the accuracy drop can be mitigated at the cost of a degradation in the channel throughput. We thus conclude that the channel remains robust in the presence of the network load provided that the size of the bursts is adjusted accordingly.

\vspace{-0.8em}
\subsection{Stealthiness}
\label{sec:stealth}
\vspace{-0.4em}

We measure the stealthiness of Bankrupt by transmitting \texttt{1}s to cause maximum memory bandwidth pressure at the intermediary (because transmitting \texttt{0}s does not generate traffic). We use Perf~\cite{perfTool} to monitor the memory traffic on the intermediary.
First, we inspect available Infiniband counters~\cite{rdma_counters}, e.g., \texttt{port\_xmit\_wait}, but none of them reveal the network round-trip delays induced by the channel activity.
Second, we use aggregate CPU counters, as, to the best of our knowledge, there are no CPUs that feature CPU counters that account for memory requests at the granularity of memory banks.

Table~\ref{bw_covertness} shows the memory traffic that is generated by an active Bankrupt channel on the intermediary. 
For the burst sizes of 32 and 128, the attack generates memory bandwidth of 0.31GB/s and 0.67GB/s, accordingly.
For modern CPUs, which feature 20-32GB/s per channel thanks to the many memory banks, the attack increases their memory bandwidth counters negligibly. 

\begin{table}[t!]
\centering
\begin{tabular}{|c|c|c|c|c|c|c|c|}
\hline
Burst      & 1024 & 512  & 256  & 128  & 64   & 32   & 16   \\ \hline\hline
Traffic & 1.04 & 0.97 & 0.82 & 0.67 & 0.48 & 0.31 & 0.18 \\ \hline
\end{tabular}
\caption{Memory traffic (in GB/s) generated by the Bankrupt channel, depending on the burst size, on the intermediary.
The maximum bank throughput on this platform is 1.22GB/s. 
\vspace{-0.4em}
}
\label{bw_covertness}
\end{table}

\begin{figure}[t!]
\centering
\includegraphics[width=\columnwidth]{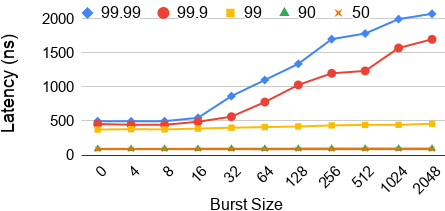}
\vspace{-2em}
\caption{Local memory access latency percentiles recorded on the intermediary during active communication over the Bankrupt channel. Burst size of 0 stands for no active channel. Lines for the 90$^{th}$ and for 50$^{th}$ percentiles appear overlapping.}
\vspace{-1em}
\label{covertness_amat}
\end{figure}

We use a random-access microbenchmark that measures the memory access latency individually using \texttt{rdtsc} register with and without the active Bankrupt channel.
Figure~\ref{covertness_amat} demonstrates the median and the tail memory latency on our cluster while the channel is not active and while the channel is active with different burst sizes. The 99$^{th}$ percentile increases by less than 10\% between the unloaded case and burst size 128. For burst size 32, the 99.9$^{th}$ and the 99.99$^{th}$ percentiles increase by approximately 20\% and 70\%, respectively. For burst sizes of 512 and larger, the 99.9$^{th}$ and 99.99$^{th}$ percentiles increase from 200\% to 400\%, but such large bursts are not necessary to transmit a message, as we show in \S\ref{isolation_performance}. 

Overall, the Bankrupt attack influences only very high percentiles of the memory access time, and injects negligible memory traffic. Capturing such subtle differences with software is challenging even in the absence of other workloads scheduled on the intermediary's CPU, or other types of the system noise, that renders the attack virtually undetectable.


\vspace{-0.4em}
\subsection{Public Cloud Performance}
\vspace{-0.4em}
Similarly to the experiments in the private cluster, we identified that the bank bits in CloudLab's servers are located at the positions of bits [6:27] in the address.

As expected for a public cloud, we found that CloudLab's network is more noisy than the private cluster in isolation (\S\ref{sec:isolation}), however less noisy than in the adversarial scenario where we inject network noise with a network loader benchmark (\S\ref{sec:net_load}). To stabilize the signal for efficient decoding in CloudLab, we use a smaller 100-bit payload size while keeping the preamble of the same 32-bit size. 

Figure~\ref{bw_acc} shows that the channel achieves 
\textasciitilde35\% less channel throughput than in the private cluster with almost the same accuracy. This bandwidth reduction can be attributed to two factors: first, the network round-trip in CloudLab is larger due to its scale, and, second, we reduced the payload size in a packet, devoting 13\% more of the channel's raw bandwidth for transmitting headers (i.e., preambles). We find that the minimal burst size that allows decoding is 32, which is equal to the minimum burst size in our experiments in the private cluster. With the burst size of 32, Bankrupt provides
the channel throughput of 74Kb/s (the accuracy is 82.2\%). Increasing the burst size to 128 allows to increase the transmission accuracy to 96.8\% while providing 51Kb/s throughput.

\vspace{-0.4em}

%% file: Sections/7-DetectionMitigation.tex
\vspace{-0.8em}

Bankrupt does not impact the local memory access time, up to the 99$^{th}$ percentile~(\S\ref{sec:stealth}). Available CPU counters account memory requests per memory controller, which is not sufficiently fine-grained to reliably detect memory traffic spikes of <1GB/s directed at just one memory bank.
We propose adding per-bank CPU counters to detect the load skew at a finer granularity. However, even though these counters are likely to reveal the attack happening, they would not allow to track down the source of the attack because many applications within the same RDMA network may have access to the remote memory on the intermediary node simultaneously.

We anticipate that architectural changes of the CPU are required to close the Bankrupt channel. As we showed in \S\ref{sec:results}, the Bankrupt channel is robust to different types of noise; as a result,  noise injection is unlikely to be an effective preventive measure. 
Instead, we suggest eliminating the root cause of the vulnerability that is the static memory interleaving scheme that RDMA exposes to an attacker. 
Hence, CPU architects may consider using a cryptographic block-cipher function, that a memory controller can use for routing memory requests to memory banks, instead of the static one. Using a cryptographic function would significantly complicate the search of addresses that belong to one bank~(\S\ref{sec:addresses}). Prior work demonstrate that a block-cipher ASIC can perform an encryption operation (e.g., for routing memory requests) in as little as 10ns, and integrating these ASICs inside memory controllers would come at the cost of <0.1\% silicon area for a modern CPU~\cite{beierle2016skinny}.
Compared to the memory access latency of a modern CPU, which is normally slightly over 100ns~\cite{7cpu}, we anticipate a moderate performance impact both for the software that runs locally on the CPU and the RDMA network latency.

%% file: Sections/8-ResponsibleDisclosure.tex
\vspace{-0.4em}

We responsibly disclosed the vulnerability to Intel on January 27th 2020, and provided our proof-of-concept code to their security team that confirmed that "an adversary can exfiltrate information leveraging the Bankrupt attack". However, Intel's response highlighted that Intel provide no security guarantees if RDMA-equipped servers are available for use of untrusted parties, i.e., in a public cloud. We disclosed our work to Microsoft Azure's security team that "determined that the attack does not pose a risk to Azure infrastructure due to their architectural decisions". 
The proof-of-concept source code is published at \url{https://github.com/ease-lab/bankrupt}.



%% file: Sections/9-RelatedWork.tex
\vspace{-0.4em}

\textbf{Single-CPU covert channels:} 
Maurice et al.~\cite{maurice2017hello} thoroughly studies the peculiarities of cache-based covert channels and demonstrates necessary communication protocol support for robust data transmission.
Wu et al.~\cite{wu2014whispers} constructs a robust covert channel between virtual machines on different cores of the same node by locking the memory bus with atomic operations. 
The DRAMA covert channel is based on the timing difference of row buffer hits and misses at DRAM memory banks that can be measured if the sender and the receiver share the same CPU~\cite{pessl2016drama}.
However, this timing difference (\textasciitilde30ns for DRAM memory) is insufficient to form a robust signal that is visible across a network.
To achieve reliable cross-network communication, Bankrupt builds up queuing inside a single memory bank with bursts of row buffer misses, elevating the timing gap to microseconds that is widely visible across an RDMA network.
Similar to this work, DRAMA reverse-engineers, using a brute-force search approach, the exact positions and functions of all the bits in a memory address but does so by running software on the target node directly. In contrast, our method allows to retrieve the positions of all the bits that identify memory banks, in a time linear in the number of address bits that is fast in practice, which is necessary to locate addresses in memory banks, across an RDMA network. Contrary to the above memory-based channels, the Bankrupt timing channel relies on inducing queuing effects in the per-bank queues inside a memory controller.

Masti et al. leverage the heat that CPU cores emit to establish covert communication~\cite{masti2015thermal}. Xiao et al. exploit memory deduplication to construct a covert channel in a virtualized environment~\cite{xiao2012covert,xiao2013security}. 
Other works rely on accesses to private~\cite{percival2005cache} and shared CPU caches~\cite{liu2015last,Maurice2015C5CC}, including the use of \texttt{clflush} instructions~\cite{gruss2016flush+}.
In contrast to all these channels, Bankrupt overcomes the requirement of colocating the sender and the receiver on the same CPU, enabling covert communication across an RDMA network. 

\textbf{Cross-network covert channels:} 
Ovadya et al. exploit the design of network protocols such as ARP, SSH, an ICMP to construct covert channels in routers in TCP/IP networks~\cite{cross-router}. Tahir et al. design a covert channel exploiting network links and routers sharing across virtual networks~\cite{tahir2016sneak}. Both of these channels deliver transmission bandwidth below 2 Kb/s in contrast to 99Kb/s provided by Bankrupt.
Recent works have discovered covert channels in modern RDMA deployments. Pythia~\cite{pythia} demonstrates a timing channel in the NIC-internal translation buffer. The channel requires the use of small (e.g., 4KB) pages for the RDMA-exposed regions while the use of large pages is widespread in settings where low latency is a priority~\cite{dragojevic:fast,guo2016rdma,novakovic19_storm,Cong2018, wang18_rdmav}.
The use of large pages immediately exposes the system to the Bankrupt attack. NetCAT designs a covert channel by leveraging Intel Data-Direct I/O (DDIO)~\cite{directtechnology} that allows buffering of RDMA packets in the destination's CPU's last-level cache (LLC)~\cite{kurth2020netcat}. Contrary to NetCAT, Bankrupt cannot be mitigated by disabling DDIO as its timing channel resides in the memory controller. 
NetCAT delivers similar throughput and accuracy as Bankrupt but lacks sufficient evaluation in the presence of noise. Bankrupt is able to deliver comparable throughput even in an noisy environment thanks to the Bankrupt channel's microsecond-scale latency gap between transmitted \texttt{0}-s and \texttt{1}-s, which is an order of magnitude larger than the LLC miss delay in NetCAT.

%% file: Sections/Conclusion.tex
\vspace{-0.4em}

Covert channels enable exfiltration of sensitive data by bypassing information containment measures from secure cloud environment to the outside world. From the attacker's perspective, an ideal covert channel should be general enough to unlock high-rate data transfers of an arbitrary size within a datacenter while remaining undetectable from a cloud vendor's monitoring capabilities. 

Our work introduces Bankrupt, a high-rate cross RDMA-network covert channel, that meets all of these requirements, by allowing the spy (sender) and the receiver malware, running on different nodes in the network, to communicate via the remote memory that is hosted on yet another -- innocuous  -- node in the same network.
We showcase that Bankrupt delivers the channel throughput of 74Kb/s
in a large-scale public cloud environment inside the Cloudlab datacenter facility, and up to 114Kb/s in the private cluster. We demonstrate that Bankrupt remains highly robust even in a noisy environment while remaining stealthy to anomaly monitoring capabilities, like Infiniband NIC and CPU counters.

%% file: Sections/Ack.tex
\section*{Acknowledgements}
\vspace{-0.4em}

The authors thank Prof. Mathias Payer for his valuable feedback on this work as well as the EASE lab members at the University of Edinburgh for the numerous discussions that gave inspiration to this work.
The research was supported by the ARM Center of Excellence at the University of Edinburgh.  